# Circulating Current Induced Electromagnetic Torque Generation in Electric Machines with Delta Windings

**Prerit Pramod**, *Senior Member*, IEEE

Control Systems Engineering, MicroVision, Inc.

*Email*: preritpramod89@gmail.com; preritp@umich.edu; prerit_pramod@microvsion.com

**Abstract**: This paper explains the phenomenon of current circulation and the resulting electromagnetic torque generation in electric machines employing delta windings. The description entails a systematic assessment of the electrical and magnetic behavior of the machine to develop mathematical models, followed by intuitive explanations of the derived analytical forms. The modeling is thoroughly validated through simulation and experimental results on a prototype machine.

## Introduction

Industrial applications requiring high control performance, such as dynamical motion control in automotive [1]–[17] and medical systems [18], expect the electric motor drives employed therein to conform to tight specification in terms of their torque output and power consumption. The design of such machines, especially where high performance and low cost are concurrent constraints, requires intricate selection of various design parameters.

Winding configurations are a key parameter in the process involved in the electromagnetic design of electrical machines due to the significant impact they have on the resulting machine performance. Moreover, the importance of winding design is ubiquitous across machine topologies, including induction, synchronous [19]–[36], switched reluctance [37]–[44], and direct current (DC) [45]–[54] machines. Delta winding configuration may be a preferred choice over its star counterpart for several applications owing to the advantages they offer including the elimination of the neutral crimp and bus bar which help in reducing the cost, improvement in the machine winding utilization, among others. The concept of delta-windings in the area of energy conversion in general [55], and on their utilization to motor design in particular [56], has been presented. More recently, hybrid winding configurations such as star-delta have been used to improve motor drive performance [57]–[59].

Delta-wound PMSMs can potentially exhibit circulating currents due to back-EMF (BEMF) voltage components which are have zero relative phase between the individual phase winding. Analysis of circulating currents has been done in the past for transformers and other electric machines, but detailed analytical models that provide an simple yet intuitive insight into the circulating currents and resulting torque has not been done extensively, especially for PMSM drives.

A generalized mathematical model for circulating currents in PMSM machines, also applicable to wound rotor synchronous machines, considering multi-phase designs, slot-pole topologies and arbitrary flux linkage harmonic content, is presented. The model is validated with simulation and experimental results on a practical PMSM drive. The model is useful for understanding circulating currents and in developing open-loop or feedforward algorithms for the compensation of the resultant torque ripple [60]–[65].

## Delta-Wound Machines

As mentioned earlier, delta-wound machines can potentially exhibit circulating currents due to back-EMF (BEMF) voltage components which are have zero relative phase between the individual phase winding. These voltage components are harmonics of orders $kn$ for a n-phase machine, where k is an integer. A simplified circuit diagram of a 3-phase delta-wound machine is illustrated in Fig. 1.

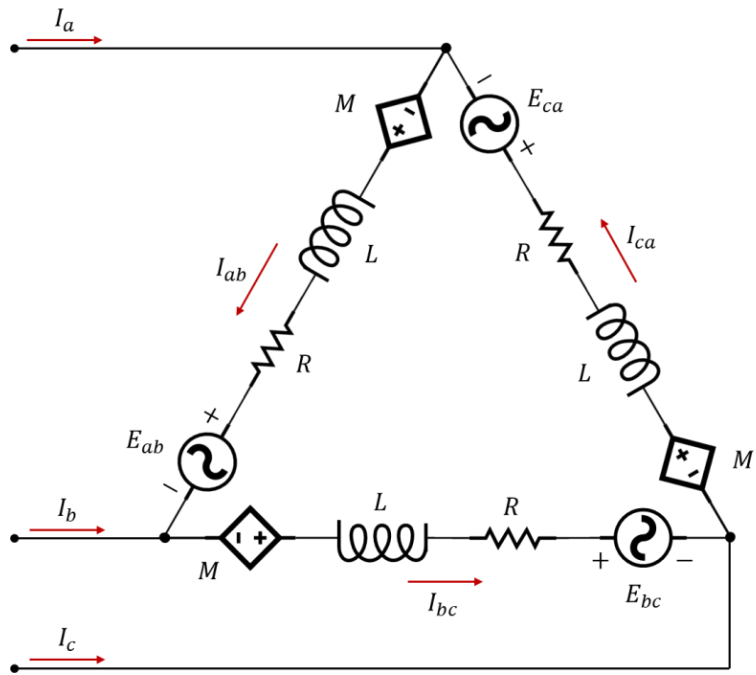

**Fig. 1**. Circuit diagram of delta-wound machine.

Notice that the parameters of the individual phases are assumed to be equal, which may not be perfectly accurate, but is typically ensured during manufacturing.

## Circulation Current & Electromagnetic Torque Modeling

Applying KVL across the winding loop results in (1).

$$E_{ab} + E_{bc} + E_{ca} + R(I_{ab} + I_{bc} + I_{ca}) + (L - 2M)(\dot{I}_{ab} + \dot{I}_{bc} + \dot{I}_{ca}) = 0 \qquad (1)$$

where $L$ and $M$ represent the self and mutual inductances, $E$ is the line-line BEMF voltage and $I$ represents the current. The sum of BEMF voltages is zero for the fundamental component and harmonics with matched magnitudes and unaligned phases, but this is not the case for $kn$ harmonics since the phases of all three voltages are identical. The general expression for the BEMF voltage of one winding may be expressed as (2).

$$E_w = E_{w|f} + \sum_h E_{w|h} \sin\big(h(\theta_e - w\beta)\big) \qquad (2)$$

where subscript $f$ refers to the fundamental component, $h$ represents the order of harmonic components, $\beta = \frac{2\pi}{n}$ for a $n$-phase machine and $w$ is an integer and may have any value between $0$ and $n - 1$. The product $hw\beta$ always results in an integer multiple of $2\pi$ when $h$ is a integral multiple of the number of phases $n$ of the machine. This in-turn causes the phases of these components to be aligned and the resultant summation is non-zero. All other harmonics add to zero if they are phased symmetrically and have identical magnitudes.

The cancellation of most of the BEMF voltage components, including the fundamental, allows for analysis of the machine in the fundamental reference frame similar to a star-wound configuration. Assuming a single harmonic component of order $h$ is dominant among the non-zero summations, the sum of the voltage components may be expressed as (3).

$$E_c = \sum_h E_h = n \, h\omega_e \, \lambda_{m|h} \sin(h\theta_e) \qquad (3)$$

where $\lambda_{m|h}$ is the magnitude of the flux linkage component of order $h$. The relationship between the flux linkage and BEMF for one winding is given as (4).

$$\lambda_w = -\lambda_{m|h} \cos(h\theta_e)$$
$$E_w = \dot{\lambda}_w = h\omega_e \lambda_{m|h} \sin(h\theta_e) \qquad (4)$$

Thus, the circulating current of order $h$ may be computed as (5).

$$I_c = -\frac{nh\omega_e \lambda_{m|h}}{\sqrt{R^2 + \big(h\omega_e(L - 2M)\big)^2}} \sin(h\theta_e - \phi_h)$$
$$\phi_h = \tan^{-1}\left(\frac{h\omega_e(L - 2M)}{R}\right) \qquad (5)$$

The electromagnetic torque generated by circulating currents may be computed as (6).

$$T_e = -p \frac{\partial}{\partial \theta_e} \sum_w \int_0^I \lambda_w \, dI_w$$

$$= pn\lambda_{m|h} \sin(h\theta_e) \, I_c$$

$$= \frac{pn^2 h \lambda_{m|h}^2}{2\sqrt{R^2 + \left(h\omega_e(L-2M)\right)^2}} (\cos\phi_h - \cos(2h - \phi_h)) \tag{6}$$

where $p$ is the number of magnetic pole pairs. Note that both (5) and (6) are obtained neglecting magnetic saturation which results in a current and position dependency of flux linkages and therefore on machine parameters. Notice that a speed dependent DC shift in average torque and a torque ripple at order $2h$ is generated as a result of circulating currents.

## Model Validation

Simulation results showing circulating currents and corresponding torque for a 3-phase 9-slot, 6-pole non-salient delta-wound PMSM are illustrated in Fig. 2.

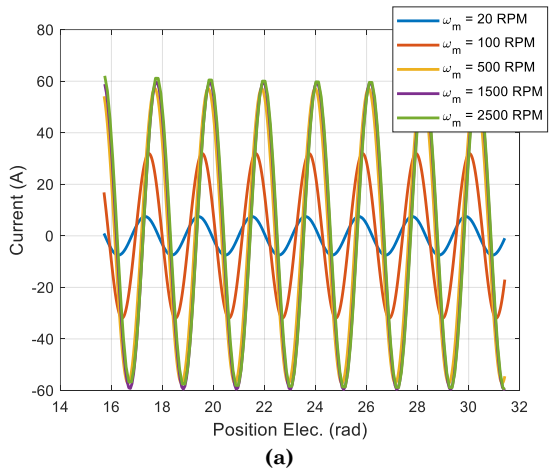

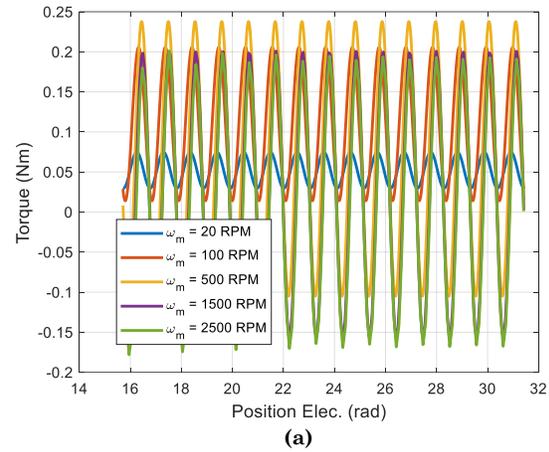

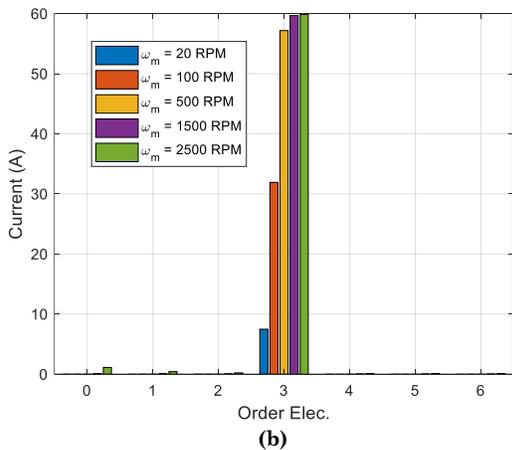

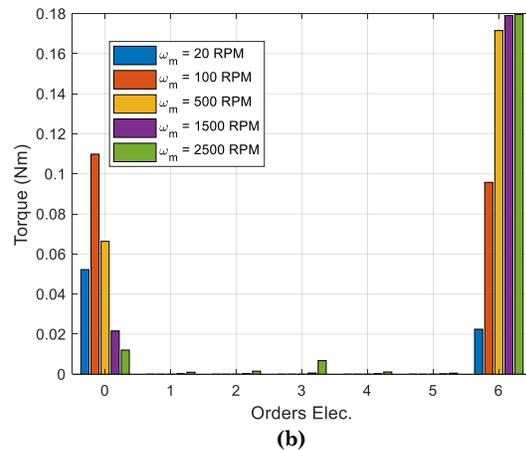

**Fig. 2**. Circulating currents (a) variation with position and (b) harmonic decomposition.

**Fig. 3**. Electromagnetic torque caused by circulating currents (a) variation with position and (b) harmonics.

The presence of $3^{\text{rd}}$ order harmonics (since dominant order $h = 3$) in circulating currents and their asymptotic dependency towards a constant value at high speeds is clear in Figs. 2(a)-(b). This is in line with the prediction in (5) where the magnitude becomes $I_c \approx \frac{n\lambda_{m|h}}{L-2M}$ at high speed. The torque variation and harmonic decomposition in Figs. 3(a)-(b) also match the analytical predictions of (6) with the DC content decaying with increasing speed due to its inverse dependency and the torque ripple at $6^{\text{th}}$ order (i.e., order $2h$) becoming constant.

Experimental back-EMF voltages and shaft torque from open-circuit tests on individual phases, star and delta configurations of an open-delta type 3-phase 12-slot, 8-pole machine are shown in Fig. 4. The individual phase results clearly show $3^{\text{rd}}$ order which are unobservable as expected in the star or delta configuration tests. Further, the difference in both the DC component and $6^{\text{th}}$ order torque outputs also aligns with the analysis presented.

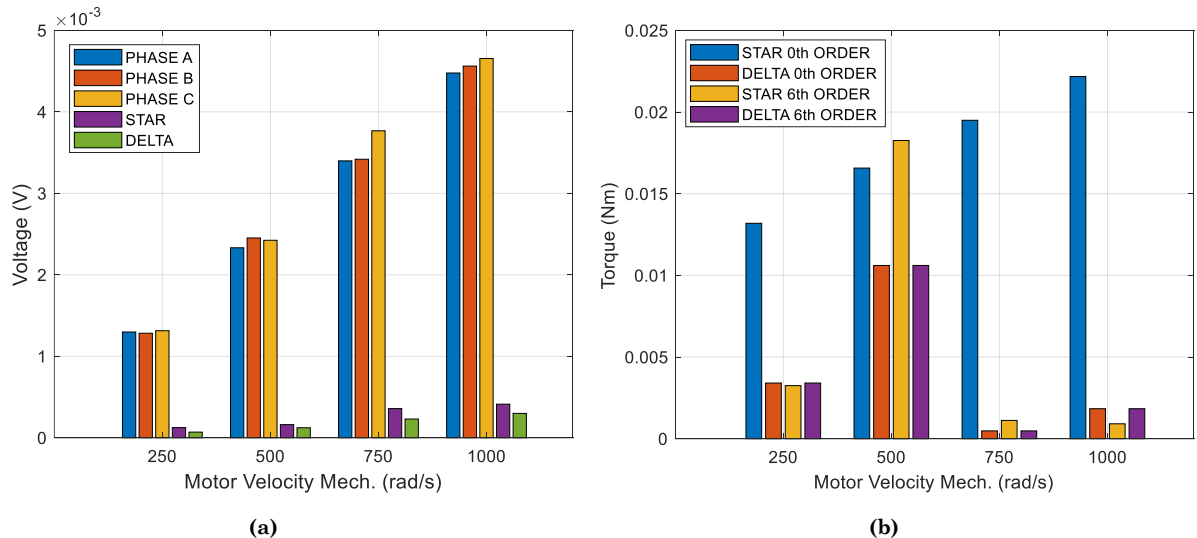

**(a)**                          **(b)**

**Fig. 4**. Experimental results for individual phases, star- and delta-connection showing harmonics of (a) back-EMF $3^{\text{rd}}$ order voltages and (b) torque electrical orders 0 and 6 for non-salient PMSM at different speeds.

## Conclusions

A generalized analytical model for circulating currents and resulting electromagnetic torque in delta wound PMSM machines is presented in this paper. In addition to the specific model derived for PMSMs, the paper provides insight into the approach required to determine similarly applicable mathematical models utilized for other machine topologies and different winding configurations. The wide utility and applicability of the modeling methodology and the derived analytical models enables its rapid deployment towards the development of associated control algorithms for motor drives.

# References

[1] Pramod, P., Mitra, R., Zuraski, J. A., Zuraski, Chandy, A., & Kleinau, J. A. (2021). *Fault tolerant field oriented control for electric power steering*. (U.S. Patent No. 10,960,922). Washington, DC: U.S. Patent and Trademark Office.

[2] Pramod, P., Shah, S. P., Kleinau, J. A., & Hales, M. K. (2018). *Decoupling current control utilizing direct plant modification in electric power steering system*. (U.S. Patent No. 10,003,285). Washington, DC: U.S. Patent and Trademark Office.

[3] Pramod, P., & Kleinau, J. A. (2021). *Current mode control utilizing plant inversion decoupling in electric power steering systems*. (U.S. Patent 11,091,193). Washington, DC: U.S. Patent and Trademark Office.

[4] Kleinau, J. A., Pramod, P., Skellenger D. B., & Sengottaiyan, S. K. (2017). *Motor control current sensor loss of assist mitigation for electric power steering*. (U.S. Patent No. 9,809,247). Washington, DC: U.S. Patent and Trademark Office.

[5] Pramod, P., & Kleinau, J. A. (2018). *Motor control anti-windup and voltage saturation design for electric power steering*. (U.S. Patent 10,103,667). Washington, DC: U.S. Patent and Trademark Office.

[6] Pramod, P., Mendon, P., Narayanaswamy, C., & Klein, F. (2021). *Impact of Electric Motor Drive Dynamics on Performance and Stability of Electric Power Steering Systems* (No. 2021-01-0932). SAE Technical Paper.

[7] Kant, N., Chitkara, R., & Pramod, P. (2021). *Modeling Rack Force for Steering Maneuvers in a Stationary Vehicle* (No. 2021-01-1287). SAE Technical Paper.

[8] Pramod, P. (2023). Control Performance Analysis of Power Steering System Electromechanical Dynamics. *arXiv preprint arXiv:2309.13623*.

[9] Wang, Q., Pramod, P., & Champagne, A. J. (2019). *Velocity estimation for electric power steering systems*. (U.S. Patent No. 10,322,746). Washington, DC: U.S. Patent and Trademark Office.

[10] Pramod, P., Mitra, R., & Ramanujam, R. (2020). *Current sensor fault mitigation for steering systems with permanent magnet DC drives*. (U.S. Patent No. 10,822,024). Washington, DC: U.S. Patent and Trademark Office.

[11] Pramod, P., Zheng, K., George, M. S., & Varunjikar, T. M. (2021). *Cascaded position control architecture for steering systems*. (U.S. Patent No. 11,203,379). Washington, DC: U.S. Patent and Trademark Office.

[12] Pramod, P., Zheng, K., George, M. S., & Varunjikar, T. M. (2021). *Disturbance feedforward compensation for position control in steering systems*. (U.S. Patent 11,180,186). Washington, DC: U.S. Patent and Trademark Office.

[13] Skellenger, D. B., & Pramod, P. (2020). *Driver warning in electric power steering systems*. (U.S. Patent No. 10,773,749). Washington, DC: U.S. Patent and Trademark Office.

[14] Pramod, P., Eickholt, M., Namburi, K. M., & Tompkins, M. A. (2021). *Dither noise management in electric power steering systems*. (U.S. Patent No. 11,117,612). Washington, DC: U.S. Patent and Trademark Office.

[15] Varunjikar, T. M., Pramod, P., Kleinau, J. A., Eickholt, M., Ramanujam, R., Skellenger, D. B., & Klein, S. D. (2020). *Driver notification using handwheel actuators in steer-by-wire systems*. (U.S. Patent No. 10,676,129). Washington, DC: U.S. Patent and Trademark Office.

[16] Naman, A., Varunjikar, T. M., Sainath, B., Ramanujam, R., & Pramod, P. (2021). *Steer by wire drift compensation*. (U.S. Patent Application No. 16/514,396).

[17] Pramod, P., Zhang, Z., Mitra, R., & Ramanujam, R. (2020). *Compensator anti-windup for motion control systems*. (U.S. Patent No. 10,773,748). Washington, DC: U.S. Patent and Trademark Office.

[18] Ozsecen, M. Y., Tosh, O. K., Mitra, R., Ryne, Z., & Pramod, P. (2019). *Assist profiling and dynamic torque generation for biomechanical assistive device*. (U.S. Patent Application No. 16/205,734).

[19] Pramod, P., Zhang, Z., Mitra, R., Paul, S., Islam, R., & Kleinau, J. (2016, June). Impact of parameter estimation errors on feedforward current control of permanent magnet synchronous motors. In *2016 IEEE Transportation Electrification Conference and Expo (ITEC)* (pp. 1-5). IEEE.

[20] Pramod, P., Zhang, Z., Namburi, K. M., Mitra, R., Paul, S., & Islam, R. (2017, May). Effects of position sensing dynamics on feedforward current control of permanent magnet synchronous machines. In *2017 IEEE International Electric Machines and Drives Conference (IEMDC)* (pp. 1-7). IEEE.

[21] Pramod, P., Zhang, Z., Namburi, K. M., Mitra, R., & Qu, D. (2018, September). Effects of position sensing dynamics on feedback current control of permanent magnet synchronous machines. In *2018 IEEE Energy Conversion Congress and Exposition (ECCE)* (pp. 3436-3441). IEEE.

[22] Pramod, P., & Namburi, K. M. (2019, September). Closed-loop current control of synchronous motor drives with position sensing harmonics. In *2019 IEEE Energy Conversion Congress and Exposition (ECCE)* (pp. 6147-6154). IEEE.

[23] Pramod, P. (2020, October). Synchronous frame current estimation inaccuracies in permanent magnet synchronous motor drives. In *2020 IEEE Energy Conversion Congress and Exposition (ECCE)* (pp. 2379-2386). IEEE.

[24] Piña, A. J., Pramod, P., Islam, R., Mitra, R., & Xu, L. (2015, September). Extended model of interior permanent magnet synchronous motors to include harmonics in d-and q-axes flux linkages. In *2015 IEEE Energy Conversion Congress and Exposition (ECCE)* (pp. 1864-1871). IEEE.

[25] Pina, A. J., Pramod, P., Islam, R., Mitra, R., & Xu, L. (2015, May). Modeling and experimental verification of torque transients in interior permanent magnet synchronous motors by including harmonics in d-and q-axes flux linkages. In *2015 IEEE International Electric Machines & Drives Conference (IEMDC)* (pp. 398-404). IEEE.

[26] Pramod, P., Govindu, V., Zhang, Z., & Namburi, K. M. P. K. (2020). *Feedforward control of permanent magnet synchronous motor drive under current sensing failure*. (U.S. Patent No. 10,717,463). Washington, DC: U.S. Patent and Trademark Office.

[27] Kolli, N., Pramod, P., & Bhattacharya, S. (2020, June). Analysis of different operating modes of PMSM during regeneration with uncontrolled rectifier. In *2020 IEEE Transportation Electrification Conference & Expo (ITEC)* (pp. 204-209). IEEE.

[28] Pramod, P., Mitra, R., Namburi, K. M., Saha, A., & Lowe, I. (2019, June). Resistance Imbalance in Feedforward Current Controlled Permanent Magnet Synchronous Motor Drives. In *2019 IEEE Transportation Electrification Conference and Expo (ITEC)* (pp. 1-5). IEEE.

[29] Pramod, P., Saha, A., Namburi, K., & Mitra, R. (2019, May). Modeling, analysis and compensation of resistance imbalance in permanent magnet synchronous motor drives for mass manufacturing applications. In *2019 IEEE International Electric Machines & Drives Conference (IEMDC)* (pp. 1106-1109). IEEE.

[30] Pramod, P., & Zhang, Z. (2019). *Controller anti-windup for permanent magnet synchronous machines*. (U.S. Patent No. 10,411,634). Washington, DC: U.S. Patent and Trademark Office.

[31] Gizinski, N. E., Pramod, P., & Kleinau, J. A. (2022). *Parameter learning for permanent magnet synchronous motor drives*. (U.S. Patent No. 11,404,984). Washington, DC: U.S. Patent and Trademark Office.

[32] Pramod, P. (2023). Performance Analysis of Synchronous Motor Drives under Concurrent Errors in Position and Current Sensing. *arXiv preprint arXiv:2310.00975*.

[33] Pramod, P. (2023). Analytical Modeling of Parameter Imbalance in Permanent Magnet Synchronous Machines. *arXiv preprint arXiv:2310.00508*.

[34] Pramod, P. (2023). Position Sensing Errors in Synchronous Motor Drives. *arXiv preprint arXiv:2310.00977*.

[35] Pramod, P. (2023). *Feedforward current control for dual wound synchronous motor drives*. (U.S. Patent No. 11,736,048). Washington, DC: U.S. Patent and Trademark Office.

[36] Pramod, P. (2023). *Current regulators for dual wound synchronous motor drives*. (U.S. Patent Application No. 17/563,908).

[37] Ma, C., Qu, L., Mitra, R., Pramod, P., & Islam, R. (2016, March). Vibration and torque ripple reduction of switched reluctance motors through current profile optimization. In *2016 IEEE Applied Power Electronics Conference and Exposition (APEC)* (pp. 3279-3285). IEEE.

[38] Ma, C., Mitra, R., Pramod, P., & Islam, R. (2017, October). Investigation of torque ripple in switched reluctance machines with errors in current and position sensing. In *2017 IEEE Energy Conversion Congress and Exposition (ECCE)* (pp. 745-751). IEEE.

[39] Mehta, S., Husain, I., & Pramod, P. (2019, September). Predictive current control of mutually coupled switched reluctance motors using net flux method. In *2019 IEEE Energy Conversion Congress and Exposition (ECCE)* (pp. 4918-4922). IEEE.

[40] Mehta, S., Pramod, P., & Husain, I. (2019, June). Analysis of dynamic current control techniques for switched reluctance motor drives for high performance applications. In *2019 IEEE Transportation Electrification Conference and Expo (ITEC)* (pp. 1-7). IEEE.

[41] Mehta, S., Kabir, M. A., Husain, I., & Pramod, P. (2020). Modeling of mutually coupled switched reluctance motors based on net flux method. *IEEE Transactions on Industry Applications*, *56*(3), 2451-2461.

[42] Mehta, S., Kabir, M. A., Pramod, P., & Husain, I. (2021). Segmented rotor mutually coupled switched reluctance machine for low torque ripple applications. *IEEE Transactions on Industry Applications*, *57*(4), 3582-3594.

[43] Pramod, P., Nuli, P., Mitra, R., & Mehta, S. (2019, May). Modeling and simulation of switched reluctance machines for control and estimation tasks. In *2019 IEEE International Electric Machines & Drives Conference (IEMDC)* (pp. 565-570). IEEE.

[44] Mehta, S., Pramod, P., Husain, I., & Kabir, M. A. (2020). Small-Signal modeling of mutually coupled switched reluctance motor. *IEEE Transactions on Industry Applications*, *57*(1), 259-271.

[45] Pandya, V., Mehta, S., & Pramod, P. (2023, May). Modeling, Characterization, and Identification of Permanent Magnet DC Motors. In *2023 IEEE International Electric Machines & Drives Conference (IEMDC)* (pp. 1-7). IEEE.

[46] Pramod, P., Mitra, R., & Ramanujam, R. (2020). *Current sensor fault mitigation for steering systems with permanent magnet DC drives*. (U.S. Patent No. 10,822,024). Washington, DC: U.S. Patent and Trademark Office.

[47] Woo, J. M., Pramod, P., & Islam, M. R. (2020). *Fault tolerant permanent magnet DC motor drives*. (U.S. Patent No. 10,756,656). Washington, DC: U.S. Patent and Trademark Office.

[48] Pramod, P. (2022). *Feedforward control of multiphase permanent magnet direct current motor drives*. (U.S. Patent No. 11,218,096). Washington, DC: U.S. Patent and Trademark Office.

[49] Pramod, P., & Govindu, V. (2019). *Disturbance observer for permanent magnet direct current machines*. (U.S. Patent No. 10,340,828). Washington, DC: U.S. Patent and Trademark Office.

[50] Pramod, P., Govindu, V., Mitra, R., & Nalakath, N. B. (2020). *Current regulators for permanent magnet DC machines*. (U.S. Patent No. 10,640,143). Washington, DC: U.S. Patent and Trademark Office.

[51] Pramod, P., & Mitra, R. (2019). *Feedforward control of permanent magnet DC motors*. (U.S. Patent No. 10,404,197). Washington, DC: U.S. Patent and Trademark Office.

[52] Pramod, P., Namburi, K. M. P., & Kleinau, J. A. (2020). *Current capability limiting of DC machines*. (U.S. Patent No. 10,530,282). Washington, DC: U.S. Patent and Trademark Office.

[53] Pramod, P., Namburi, K. M. P., & Kleinau, J. A. (2022). *Supply current limiting of DC machines*. (U.S. Patent No. 11,290,042). Washington, DC: U.S. Patent and Trademark Office.

[54] Pramod, P., Namburi, K. M. (2020). Regenerative current limiting of DC machines. (U.S. Patent No. 10,618,547). Washington, DC: U.S. Patent and Trademark Office.

[55] Faccioli, G. (1922). Triple harmonics in transformers. *Journal of the American Institute of Electrical Engineers*, *41*(5), 351-359.

[56] Kallesoe, C. S., Izadi-Zamanabadi, R., Vadstrup, P., & Rasmussen, H. (2007). Observer-based estimation of stator-winding faults in delta-connected induction motors: A linear matrix inequality approach. *IEEE Transactions on Industry Applications*, *43*(4), 1022-1031.

[57] Sadeghi, S., Guo, L., Toliyat, H. A., & Parsa, L. (2011). Wide operational speed range of five-phase permanent magnet machines by using different stator winding configurations. *IEEE Transactions on Industrial Electronics*, *59*(6), 2621-2631.

[58] Vidmar, G., & Miljavec, D. (2014). A universal high-frequency three-phase electric-motor model suitable for the delta- and star-winding connections. *IEEE transactions on power electronics*, *30*(8), 4365-4376.

[59] Ibrahim, M. N., Sergeant, P., & Rashad, E. E. M. (2016). Combined star-delta windings to improve synchronous reluctance motor performance. *IEEE Transactions on Energy Conversion*, *31*(4), 1479-1487.

[60] Pramod, P., Lowe, I., Namburi, K. M., & Govindu, V. (2019). *Torque ripple compensation with feedforward control in motor control systems*. (U.S. Patent No. 10,333,445). Washington, DC: U.S. Patent and Trademark Office.

[61] Pramod, P., & Kleinau, J. A. (2022). *Torque ripple compensation in motor control systems*. (U.S. Patent No. 11,515,813). Washington, DC: U.S. Patent and Trademark Office.

[62] Pramod, P. (2022). *Open loop compensation of current harmonics induced by parameter imbalance*. (U.S. Patent No. 11,411,515). Washington, DC: U.S. Patent and Trademark Office.

[63] Ballal, S., Pramod, P., & Kleinau, J. A. (2018). *Torque ripple cancellation algorithm involving supply voltage limit constraint*. (U.S. Patent No. 10,135,368). Washington, DC: U.S. Patent and Trademark Office.

[64] Pramod, P., Zhang, Z., Namburi, K. M., Lowe, I. O., & Saha, A. (2022). *Closed-loop compensation of current measurement offset errors in alternating current motor drives*. (U.S. Patent No. 11,378,597). Washington, DC: U.S. Patent and Trademark Office.

[65] Pramod, P. (2022). *Feedback compensation of parameter imbalance induced current harmonics in synchronous motor drives*. (U.S. Patent No. 11,349,416). Washington, DC: U.S. Patent and Trademark Office.